\documentclass[conference,twocolumn]{IEEEtran}
\usepackage{graphicx}
\usepackage{amsmath, amssymb}
\usepackage{float}
\usepackage{algorithmic}

\IEEEoverridecommandlockouts
\newtheorem{theorem}{Theorem}
\newtheorem{lemma}{Lemma}
\newtheorem{proposition}{Proposition}
\floatstyle{ruled}
\newfloat{algorithm}{tbp}{loa}
\floatname{algorithm}{Algorithm}

\title{Power Aware Routing for Sensor Databases}
\author{
\authorblockN{Chiranjeeb Buragohain,  Divyakant Agrawal, and
Subhash Suri\thanks{The authors were supported by 
NSF  grants IIS-0121562,  CNS-0423336 and Army Research Organization
grant DAAD19-03-D-0004 through the Institute for Collaborative 
Biotechnologies.}}
\authorblockA{
Dept. of Computer Science\\ University of California, Santa Barbara, 
CA 93106\\
\texttt{\{chiran,agrawal,suri\}@cs.ucsb.edu}}
}

\date{}
\begin{document}
\maketitle
\begin{abstract}
Wireless sensor networks offer the potential to span and monitor large
geographical areas inexpensively.  Sensor network databases like
TinyDB \cite{madden02supporting} are the dominant architectures to
extract and manage data in such networks.  Since sensors have
significant power constraints (battery life), and high communication
costs, design of energy efficient communication algorithms is of great
importance.  The data flow in a sensor database is very different from
data flow in an ordinary network and poses novel challenges in
designing efficient routing algorithms.  In this work we explore the
problem of energy efficient routing for various different types of
database queries and show that in general, this problem is NP-complete.
We give a constant factor approximation algorithm for one class of
query, and for other queries give heuristic algorithms.  We evaluate
the efficiency of the proposed algorithms by simulation and
demonstrate their near optimal performance for various network sizes.
\end{abstract}

\section{Introduction}

We consider the problem of energy efficient routing in a sensor
network for data aggregation queries.  In such networks, energy
costs are dominated by communication and we give algorithms to
optimally route aggregated data through the network.
\subsection{Motivation}
\emph{Sensor networks} are large scale wireless networks composed of
tiny \emph{sensor nodes}.  Each sensor node is a two part combination
of an array of sensors and a tiny computer capable of wireless
communication.  The sensors periodically measure the physical
environment around them and generate a stream of data.  The computer
takes this data and communicates it through a wireless network to
other nodes in the network.  Sensor database systems like TinyDB
\cite{madden02supporting} have been proposed 
to provide a SQL like data
interface to simplify the extraction and management of data in such a
network.

In the Internet, data extraction and routing are relegated to
different layers in the protocol stack.  But in a sensor network, such
separation of application and network layers is not desirable.  Most
sensor network are \emph{not} designed to be general purpose networks
and thus their architecture can be fine tuned for the intended
application.  Most importantly, sensor networks are severely
power-limited: sensor nodes run on battery power, which is generally
impossible to replenish once the network is deployed.  Thus the
lifetime and usefulness of the network crucially depends on the energy
demands that the application places on it.  A node can spend energy in
sensing, communication and computation.  In typical sensors,
computation and sensing costs are several orders of magnitude smaller
than the communication cost.  As an example, for Berkeley sensor
motes, transmitting one bit requires 1 $\mu$J, while executing one
instruction takes about 0.008 $\mu$J \cite{hill00system} of energy.
Thus the biggest power savings can be achieved by communication
strategies which are optimal for the application at hand.


\subsection{Routing Challenges in Sensor Database Applications}
In a sensor database system, users pose queries to a special node
called the \emph{base station} which disseminate the queries over the
network.  An example query could be to find the average temperature
over all sensors in the network.  Sensors process the query and send
their replies back to the base station.  The most natural way to
answer queries is to gather data from all the sensors at the base
station and process the query there.  Since this process is very
expensive in terms of the amount of data transferred, systems such as
TinyDB have proposed \emph{in-network aggregation} to process queries
~\cite{madden02tag}.  In-network aggregation has resulted in routing
trees playing a central role in the database application.  Prior
research on power aware routing \cite{chang00energy,
sadagopan04maximizing, zussman03energy} and query processing has
considered the two problems in isolation.  In particular, because of
in-network aggregation many data packets are often aggregated to a
single packet.  The typical energy efficient routing schemes do not
take into account this collapse of data flow.  Similarly database
query processing techniques have critically relied on the availability
of a routing tree for in-network aggregation without considering the
optimality of such a tree.

In this paper we first classify database queries based on the level of
aggregation they employ for processing.  If a query can be processed
in-network by transmitting a single value from each sensor node on the
routing tree, we refer to this as a \emph{fully-aggregated query}.  The other
case is when an intermediate node needs to transmit all incoming
values toward the base station; this is referred to as an 
\emph{unaggregated query}.  The final category which we refer to as 
\emph{partially aggregated query} is one in which the amount of data 
each node needs to transmit
has an upper bound, e.g. a histogram over the data values.  Each type
of in-network aggregation results in a different optimal routing tree.
Although transmission costs dominate wireless communication, reception
costs also play a strong role in determining optimal routing tree.
For example, the radio for MICA2 sensor motes drain 12mA current while
transmitting and 8mA while receiving \cite{crossbow03MPR}.
Simply keeping the radio switched on at idle also drains
consumes almost the same amount of power as receiving.  Thus both
receive and idle time in a network can significantly affect the total
energy consumption.

The problem of finding power efficient routing tree arises in other
contexts as well.  For example some sensor network routing schemes
divide sensors into clusters with every node reporting to its cluster
leader.  The nodes in a cluster can transmit to the cluster leader in
a single hop, or use lower power and try to reach the leader in
multiple hops.  Multiple hops are generally more power efficient than
single hops and in that case we have a similar problem of needing to
find the best routing tree to reach the cluster leader.

\subsection{Our Contribution}
We show that finding the optimal routing tree is NP-complete for aggregated, 
unaggregated or partially aggregated queries.
For fully aggregated
queries we give a routing algorithm with worst case constant factor
performance guarantee.  For unaggregated and partially aggregated
queries we give approximation algorithms which in practice come very
close to optimal performance.  Finally we evaluate these algorithms
experimentally in various network topologies.

The rest of the paper is organized as follows.  In Section
\ref{sec-background} we discuss the properties of our model and
discuss some related work.  Then in Sections
\ref{sec-fully-aggregated}, \ref{sec-unaggregated} and
\ref{sec-partly-aggregated}, we discuss routing problems for fully
aggregated, unaggregated and partially aggregated queries.  Section
\ref{sec-experimental} is devoted to an experimental evaluation of the
proposed algorithms  and section \ref{sec-conclusion} concludes the paper.

\section{Background and Related Work}
\label{sec-background}
\subsection{The Model}
Consider a network of $N$ sensor nodes labeled as $i, i = 1,2,\ldots
N$ and having energy $e_i, i = 1,\ldots, N$.  We assume that the node
$1$ acts as the base station.  This node is special in that it is
assumed to have unlimited power supply, i.e. $e_1 = \infty$.
We assume a setting in which the environment is continually monitored by
the sensor and the data value is sent to the base station.  Formally,
we assume that every node generates one unit of data every time unit.
Transmitting one unit of data costs one unit of energy, while
receiving one unit of data costs $c_r < 1$ amount of energy for each
sensor.

Let us assume that the $N$ sensors are distributed arbitrarily in a two 
dimensional plane and that that two sensors can communicate with each other if
they are within range $r$. We assume that sensors transmit with fixed power; 
that is, they do not dynamically adjust their power for each transmission, 
depending on the distance to the receiver node.
The radio links are also assumed to be symmetric.  In that case, we can
construct a communication graph where each node corresponds to a
sensor and each edge links two sensors which can communicate with each
other.

We assume that there exists a time synchronization model for computing
aggregates.  Time is divided into equal sized periods called
\emph{epochs}.  Every epoch a leaf node transmits data to its parent.
Once a parent receives data from all its children, it aggregates the
data and transmits the aggregate to its parent in the next epoch.
Thus if the depth of the routing tree is $d$, then it requires $d$
epochs for the data from a leaf to reach the base station.  Our task is
to construct a routing tree on this graph so as to optimize system
lifetime.  Note that in any routing tree the nodes near the base
station will be the most loaded because they have to route all the
information from other nodes to the base.

Let us now consider how a query is answered using in-network
aggregation within the Tiny AGgregation (TAG) \cite{madden02tag}
framework.  Consider the query \texttt{SELECT AVG(temp) FROM sensors}.
The base station floods this query to the network.  As the query is
propagated through the network the nodes organize themselves into a
\emph{routing tree} with the base station as the \emph{root}.  When a
node hears a query it forwards the query to its children.  When the
nodes begin to reply, every node aggregates the replies from its
children and forwards the aggregated reply to its parent.  The data
structure which encodes the reply is called a \emph{partial state
record} (PSR).  For the AVG query, the PSR consists of a tuple
$\langle \texttt{sum, count} \rangle$.  For the leaf nodes,
\texttt{count} is 1 and \texttt{sum} is the temperature reading from
the sensor.  For a parent node which receives two PSRs $\langle
\texttt{sum1,count1} \rangle$ and $\langle \texttt{sum2,count2}
\rangle$, the new PSR is
\begin{eqnarray}
  \langle \texttt{sum,count}\rangle&=&\langle \texttt{sum1+sum2+
  sensor\_value,} \nonumber\\
  & & \texttt{count1+count2+1}\rangle \nonumber
\end{eqnarray}

For a sensor network the power consumption is determined by the size
of the PSR and the path that the PSR takes to reach the root.  Here we
summarize some example queries and associated PSR sizes.
    \begin{itemize}
      \item \emph{Fully Aggregated}: \texttt{SELECT AVG(temp) FROM
      sensors}.  The PSR size is constant irrespective of the number
      of sensors.  \texttt{MAX/MIN} queries also fall into this class.
      \item \emph{Unaggregated}: \texttt{SELECT nodeid FROM
      sensors WHERE (temp $>$ 70)}.  The PSR size near the root is
      proportional to the number of total sensors.
      \item \emph{Partially Aggregated}: \texttt{SELECT HISTOGRAM(temp) FROM
      sensors}.  The PSR grows in size as it flows toward the root,
      but has a maximum size (number of bins in the histogram) independent
      of the number of sensors.
    \end{itemize}
Several metrics for power efficiency have been proposed in the
literature \cite{singh98poweraware,chang00energy} such as total energy
spent, energy spent per unit of data transmitted, and time for first
node failure.  Intuitively, an efficient routing tree will steer most
of the traffic away from the nodes with low battery power and put it
through the nodes which have more battery power and thus make sure
that no node runs out of power prematurely.  The success of the
routing tree will be determined by the length of time it can avoid a
node failure.  Thus we define a metric called \emph{system lifetime}
which is the time required for first node failure to occur amongst all
nodes.  The problem that we address in this paper is as follows: for a
given type of query with associated PSR size and a single base
station, what is the optimal routing tree to maximize system lifetime?


\subsection{Related Work}
A description of the TinyDB system and its architecture can be found
in the papers by Madden et. al. \cite{madden02tag,
madden02supporting}.  An overview of the energy consumption issues for
sensor networks can be found in the review article by Raghunathan et.
al.~\cite{raghunathan02energy}.  


The work by Singh et. al.~\cite{singh98poweraware} introduced various
metrics for measuring the lifetime of wireless networks.  
Chang and Tassiulas~ \cite{chang00energy} formulated the problem of
routing multiple conserved data flows as a multi-commodity flow with
the time for first node failure as the power efficiency metric.
Sadagopan and Krishnamachari~\cite{sadagopan04maximizing} have
proposed another efficiency metric which is to maximize total amount
of data extracted out of the network.  This is not a suitable metric
for sensor databases because the usefulness of a sensor database
depends not only on the total amount of data extracted, but also on
having as many of the nodes alive as possible.  Zussman and 
Segall~\cite{zussman03energy} have investigated power aware routing with
first node failure as efficiency metric for disaster recovery
networks.

Kalpakis et al. ~\cite{kalpakis02maximum} have formulated the power
aware routing problem for fully aggregated queries like AVERAGE as a
linear programming problem.  Krishnamachari
et.al.~\cite{krishnamachari02impact} have looked at the power
efficient routing tree problem for fully aggregated data with power
metric as total number of transmissions required.  This problem maps
on to the well known Steiner tree problem which is known to be 
NP-complete.  However, the total number of transmissions is not a good
measure of useful system lifetime.  
Boukerche et. al.~\cite{boukerche03energy} have proposed an alternative way to conserve
energy, which is to build a routing tree with a large number of
leaves.  The leaf nodes can turn themselves off to sleep until an
interesting event occurs, while the non-leaf nodes are always awake
for routing.  The problem of determining spanning tree with maximum
leaves is again known to be NP-complete.  This approach is not
applicable to database queries where all nodes need to be queried.
Also a disproportionate burden of power is placed on backbone non-leaf
nodes.  Another way to increase power efficiency of a network is to
reduce path length by utilizing multiple base station nodes.  Bogdanov
et al.~\cite{bogdanov04poweraware} have demonstrated that optimal
power efficient placement of base stations is NP hard.


\section{Routing Tree for  Fully Aggregated Queries}
\label{sec-fully-aggregated}
Let us consider the following model for a fully aggregated query.  At
every unit of time, every node produces one unit of data.  A node
which receives multiple units of data from other nodes aggregates that
data with its own and forwards a single unit of aggregated data.  Thus
every node transmits a single unit of data, but might receive zero or
more units of data depending on network topology and routing.  An
example of fully aggregated query is computing the average of data
values over all sensors.

In this section we first discuss the problem of finding optimal
routing tree for fully aggregated queries with and without reception
cost.  We demonstrate that the problem of finding optimal routing
tree with reception cost is NP-complete.  Then we give a near optimal
routing tree algorithm for the problem and analyze its performance.

If we assume that there is no energy cost to receive data, then
finding the optimal routing tree is a simple problem.  The energy cost
for any node is just the cost of transmitting one unit of data to its
parent in the routing tree regardless of its energy level.  Thus in
any routing tree, every node consumes equal amount of energy
regardless of the topology.  To state it formally:

\begin{proposition}
\label{prop-optimal-tree}
In the model where the receive cost is assumed to be zero, every 
spanning tree is optimal for aggregated queries, even with arbitrary
node energy levels.
\end{proposition}

In the introduction we have already seen that reception cost is only
marginally smaller than transmission cost for sensor nodes.  In fact,
for fully aggregated queries, a node will receive packets from
multiple neighbors, but transmit to only one parent node.  Thus, for
fully aggregated queries, the total reception cost can be significantly
larger than the transmission cost.

Surprisingly, it turns out that once we introduce reception cost into
the model, the problem of finding optimal routing tree becomes
intractable.  The amount of data received by a node is dependent on
the number of neighbors and thus to minimize power consumption a node
should have as few neighbors as possible.  This optimization problem
is NP-complete, as shown by the following theorem.


\begin{theorem}
Finding maximum lifetime routing tree for fully aggregated queries
with reception costs is NP-complete.
\end{theorem}
\begin{proof}
We shall show that the problem is NP-complete even when all nodes have
equal power.  The total energy consumption for node $i$ per unit time
is $1 + (\textrm{nbr}(i)-1)c_r$ where $c_r$ is the cost to receive one
unit of data and nbr($i$) is the number of neighbors of node $i$ in
the routing tree.  Thus to maximize lifetime we need to find a
spanning tree for the graph such that the number of neighbors for each
node is the minimum, or equivalently we need to minimize the maximum
degree of the spanning tree.  But this problem is the same as
\textsc{Minimum Degree Spanning Tree} (MDST) which is known to be
NP-complete~\cite{garey79computers}.

\end{proof}

At this point we would like to point out an upper bound on the optimal
routing lifetime $T_\textrm{OPT}$.  Consider a network where the
minimum energy node has energy $e_\textrm{min}$.  The energy
consumption for this node is minimized when it acts as a leaf in the
routing tree and does not need to receive any data.  In that case the
lifetime of this node itself is $e_\textrm{min}/1 = e_\textrm{min}$
(recall that transmission cost is 1).  Thus we have the following
upper bound:
\begin{equation}
  T_\textrm{OPT} \leq e_\textrm{min}.
\label{eqn-upper-bound}
\end{equation}

\subsection{A Near Optimal Routing Tree for Fully Aggregated Queries}
Since finding optimal routing tree is NP-complete, in this section we
present an algorithm to find a near optimal routing tree.  To solve
this problem, we convert the optimization problem to a decision
problem.  In other words, let us ask the following question: given a
network topology and node energy levels, is there a routing tree which
has a lifetime $T$?  Then the optimization problem which corresponds
to maximizing $T$ can be solved by performing a binary search on $T$.

We note that for \emph{unequal energy levels} for nodes, MDST is not
the optimal solution to the routing tree problem.  A node with large
energy can support more neighbors than a node with little energy.  So
our solution strategy is two step.  In the first step we reduce the
general routing tree problem to the MDST problem.  In the second step
we solve the MDST problem using an approximation algorithm given by
F\"{u}rer and Raghavachari\cite{furer92approximating,
raghavachari97algorithms} in the context of general spanning tree
algorithms.

\subsubsection*{Transformation of Routing Problem to the MDST Problem}
Let us consider a network with nodes having unequal energies $e_i$.
If the cost of receiving one unit of data is $c_r$ ($0 < c_r < 1$),
then to achieve a lifetime of $T$, each node $i$ can have at most
$B_i$ neighbors such that the following power constraint is satisfied:
\begin{equation}
  \frac{e_i}{1+c_r(B_i-1)} \geq T,
\label{eqn-mdst-time}
\end{equation}
which means that the maximum number of neighbors that a node $i$ can
have is
\begin{equation}
  B_i = \left\lfloor 1+\frac{1}{c_r}\left(\frac{e_i}{T}-1\right)\right\rfloor.
\label{eqn-mdst-limit}
\end{equation}
For the root node there is no power constraint and hence we set
$B_{\mathrm{root}} = N$.  Now, for every node $i$ in the original
graph, introduce $N-B_i$ auxiliary nodes and connect them to node $i$.
We call this graph with auxiliary nodes the \emph{augmented graph}.
We claim that if we construct the MDST on this graph and from the
resulting MDST delete the auxiliary nodes, the resulting spanning tree
will have degree at most $B_i$ for every node $i$.

To prove this claim, consider the original and the augmented graphs
and their respective spanning trees.  Let us call the optimal spanning
tree with non-uniform degree bounds $B_i$ to be the Non-uniform
Minimum Degree Spanning Tree (NMDST).  Suppose there exists an NMDST
on the original graph which satisfies the condition $\mbox{deg}(i)
\leq B_i$ for all $i$ where $\mbox{deg}(i)$ is the degree of node $i$
in the NMDST.  The MDST on the augmented graph is simply the NMDST
with the auxiliary nodes attached to it.  Then every node $i$ on the
MDST satisfies the \emph{uniform} degree bound $\mbox{deg}(i) \leq N$
by construction (we have introduced $N-B_i$ auxiliary nodes for each
node $i$).  Conversely, we see that if there exists an MDST with
degree bound $N$, then there exists an NMDST which satisfies the
degree bound $\textrm{deg}(i) \leq B_i$.

\begin{figure}
  \includegraphics[width=0.45\textwidth]{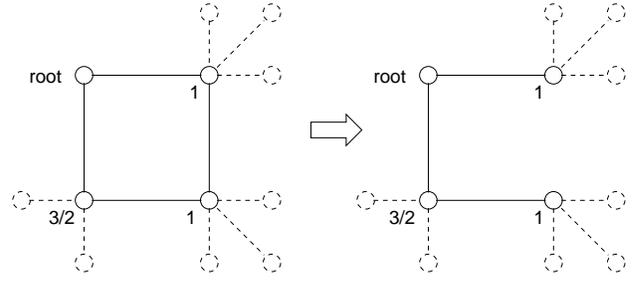}
\caption{Finding \textsc{Non-uniform Minimum Degree Spanning Tree}
  with node degree bounds.  On the left is the augmented graph, while
  on the right we have the resulting NMDST.  The auxiliary nodes are
  shown in dashed line.}
\label{fig-mdst}
\end{figure}

We illustrate graph augmentation with the example shown in
Fig. \ref{fig-mdst}.  We have a 4 node graph whose nodes have energy
levels 1 and 3/2.  The actual nodes and edges in the graph are shown
in solid lines, while auxiliary nodes and auxiliary edges are shown in
dashed lines.  With $c_r = 1/2$ and $T=1$ we find that we must
introduce 3 auxiliary nodes for the nodes with energy 1 and 2
auxiliary nodes for the node with energy 3/2.  Finding the MDST on
this graph and deleting the auxiliary nodes immediately leads to the
correct routing tree with lifetime 1.

\subsubsection*{Solving the MDST Problem}
Now we can turn our attention to the MDST problem itself.  The MDST
problem is known to be approximable to within \textsc{OPT+1}
\cite{furer92approximating, raghavachari97algorithms}.  In other
words, if the maximum degree of minimum degree spanning tree is
\textsc{OPT}, then there is a polynomial time algorithm which can
produce a spanning tree with maximum node degree \textsc{OPT+1}.  It
is easy to convince oneself that if the MDST problem can be solved
within OPT+1, then the NMDST problem also can be solved such that
$\textrm{deg}(i) \leq B_i+1$.  A detailed discussion of the OPT+1
approximation algorithm for finding the MDST can be found in the paper
by F\"{u}rer and Raghavachari \cite{furer92approximating,
raghavachari97algorithms}.  Here we describe the algorithm without
justifying it.

The full algorithm to find the routing tree is described here as the
algorithm AGGREGATED-TREE (algorithm \ref{algo-aggregated-tree}).
The algorithm takes as input a graph $G(V,E)$ with vertices $V$ and
edges $E$ and outputs a tree.  The algorithm runs in
$\mathcal{O}(NE\alpha(E, N)\log N)$ time
\cite{raghavachari97algorithms} where $\alpha$ is the inverse
Ackermann function, $E$ is the number of edges.

\begin{algorithm}
\caption{AGGREGATED-TREE($G(V, E), e$)}
\label{algo-aggregated-tree}
\begin{algorithmic}[1]
\STATE For each vertex $v$, augment it with $N-B_v$ vertices according
to eqn. \ref{eqn-mdst-limit}

\STATE Find a spanning tree $\mathcal{T}$ of $G$.  Let $k$ be its maximum degree

\STATE Mark all vertices of degree $k$ and $k-1$ as bad. Mark all
other vertices as good.

\STATE Remove all bad vertices of degree $k$ and $k-1$ generating a
forest $F$ from $\mathcal{T}$.

\WHILE{there is an edge $(u,v)$ connecting two different components in $F$ 
and all vertices of degree $k$ are marked bad}

\STATE Find the cycle $C$ generated by $\mathcal{T}$ and the edge $(u,v)$

\STATE  Mark all bad vertices in $C$ as good

\STATE Update $F$ by combining the components and vertices along cycle $C$.

\ENDWHILE
\COMMENT {At this point we have identified a vertex of degree $k$
which is in a cycle $C^\prime$}

\IF {there is a vertex $w$ of degree $k$ marked good}

\STATE Delete an edge incident on $w$ to break the cycle $C^\prime$
and reduce the degree of $w$.

\STATE Update $\mathcal{T}$ and if necessary update $k$.

\STATE Goto 3.

\ENDIF 

\STATE Delete all auxiliary nodes from $\mathcal{T}$ and output  $\mathcal{T}$.
\end{algorithmic}
\end{algorithm}

Let us denote the lifetime achieved by the optimal tree as
$T_\textrm{OPT}$ and lifetime achieved by the approximation as
$T_\textrm{MDST}$.  Then using eqn. \ref{eqn-mdst-time}
\begin{equation}
  \frac{T_\textrm{OPT}}{T_\textrm{MDST}} = \frac{1+c_rB_i}{1+c_r(B_i-1)} 
  = 1 + \frac{c_r}{e_i}T_\textrm{OPT} \nonumber
\end{equation}
Since the minimum value of $e_i$ is $e_\textrm{min}$, and
$T_\textrm{OPT} \leq e_\textrm{min}$ (eqn. \ref{eqn-upper-bound}),
\begin{displaymath}
  \frac{T_\textrm{OPT}}{T_\textrm{MDST}} \leq
  1+\frac{c_r}{e_\textrm{min}}T_\textrm{OPT}
  \leq 1+c_r < 2
\end{displaymath}
Thus our approximation scheme is within a constant factor $(1+c_r)$ of
the optimal routing tree.  As expected from proposition
\ref{prop-optimal-tree}, for zero receive cost, the approximate
lifetime coincides with the optimal lifetime.
\section{Routing Tree for Unaggregated Queries}
\label{sec-unaggregated}
In unaggregated queries the volume of data is conserved.  All data
that is generated in the nodes must be delivered to the root.  An
example of such a query is \texttt{SELECT nodeid FROM sensors WHERE
(temp $>$ 70)}.  If we assume that the condition \texttt{(temp $>$
70)} is satisfied with probability $p_t$ by the sensors, then on the
average, every sensor node produces $p_t$ units of data every time
period.  Thus the problem is to route a total of $Np_t$ data every
time unit from the nodes to the root with every node generating $p_t$
data and no data getting lost.  With a rescaling of energy or time
unit, we can convert it to the problem where every node generates one
unit of data.

Unlike the fully aggregated routing tree problem, finding optimal
routing tree for this problem is NP-complete \emph{even when we do not
take into account the cost to receive data} (Theorem
\ref{thm-unaggr}).  So for the sake of simplicity, in this section we
shall take account of only transmission costs.  But as the following
argument shows, ignoring reception cost does not invalidate our
conclusions for realistic networks.  In the optimal routing tree,
consider the node which is most likely to run out of power first.  The
data inflow and outflow for this node are $f_\textrm{in}$ and
$f_\textrm{out} = f_\textrm{in}+1$ respectively.  Total energy
consumption per unit time for this node is
\begin{displaymath}
f_\textrm{out}+c_r(f_\textrm{out}-1) = 
f_\textrm{out}(1+c_r) -c_r
\end{displaymath}
Since this is the highest loaded node, $f_\textrm{out} \gg 1 > 
c_r$, and we can say that total power consumption is proportional to the
transmission power consumption.  Thus by rescaling the energy of the
nodes by a factor of $\frac{1}{1+c_r}$, we can take into account the
most important effects of reception cost.  The plan for the rest of
the section is as follows: we first show that the problem of finding
optimal routing tree without reception cost is NP-complete.  Then we
formulate the routing tree problem as an integer program and derive an
upper bound on the maximum lifetime.  Next we give two approximation
algorithms to solve this problem and discuss their properties.

\subsection{Hardness of the Optimal Routing Tree Problem}
To establish the hardness of the optimal routing tree problem, we
recast it as a decision problem: given a connectivity graph and node
energies, does there exist a routing tree with a given lifetime $T$?
\begin{theorem}
Finding maximum lifetime routing tree for unaggregated queries is 
NP-complete.
\label{thm-unaggr}
\end{theorem}
\begin{figure}
  \includegraphics[width=0.45\textwidth]{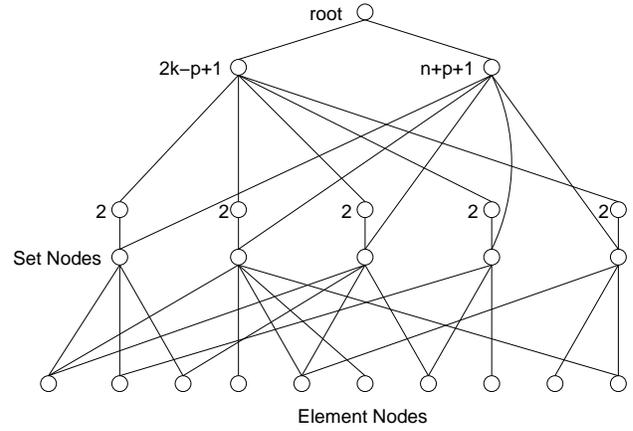}
  \caption{Reduction from \textsc{Set-Cover} to the unaggregated
  routing tree problem.  The numbers next to a node indicates the
  energy level of the node.  The set nodes have energy equal to the
  size of the set + 1.  The element nodes have energy 1.}
  \label{fig-reduction}
\end{figure}
\begin{proof}
The decision problem for this instance is clearly in NP.  Given a
graph and the routing tree on it, it is easy to verify whether the
routing tree achieves the lifetime $T$.  To show NP-completeness, we
shall exhibit a reduction from \textsc{Set-Cover} to the current
problem.  The decision problem for \textsc{Set-Cover} is as follows:
we are given $n$ elements numbered from $1\ldots n$.  We are also
given $k$ subsets of these elements ${S_1, S_2, \ldots, S_k}$.  The
problem is to decide if there exists a selection of $p$ subsets from
the collection of subsets such that the union of $p$ subsets cover all
the $n$ elements.

Given this instance of \textsc{Set-Cover}, we construct the following
instance of the communication graph.  The nodes in the graph are
arranged in five rows as shown in Fig. \ref{fig-reduction}.  The first
row consists of the root node.  The second row consists of two nodes,
one with energy $2k-p+1$ and the other with $p+n+1$.  Both these nodes
are connected to the root node.  These two nodes are used to decide
which sets will be in the set cover and which will not.  The third row
consists of $k$ nodes each with energy 2.  All the nodes in the third
row are connected to the node with power $2k-p+1$.  The fourth row
consists of $k$ nodes which corresponds to the $k$ subsets.  The
$i-$th node in the row has energy $|S_i|+1$.  Each of these nodes are
connected to the corresponding node on the third row.  Also each of
these nodes is connected to the node on second row with energy
$p+n+1$.  The fifth and last row corresponds to the $n$ elements and
each node has unit energy.  Each node corresponding to $S_i$ is
connected to the elements that it contains.  For example if $S_2$
contains the elements $(2,5,6,10)$, then those nodes have links to the
node corresponding to $S_2$.

Now we claim that there exists a routing tree with lifetime 1 if and
only if a set cover of size $p$ exists.  Suppose there exists a set
cover of size $p$.  Then we can construct the routing tree as follows.
The data from the $p$ sets and the $n$ elements can be routed through
the node with power $n+p+1$ to the root.  The data from rest of the
$k-p$ subsets and the $k$ nodes in the third row can be routed through
the node with power $2k-p+1$.  Conversely if there is a routing tree
of lifetime 1, then the third row nodes with energy 2 ensure that they
are carrying data from only $k-p$ subsets.  Thus the rest of the $p$
nodes constitute a set cover.
\end{proof}
Although we have used nodes with different powers in this proof, it
can be shown \cite{buragohain04power} that this problem remains
NP-complete even if we constrain all nodes to have the same
energy.  The proof, which we omit, relies on a pseudo-polynomial
reduction \cite{garey79computers} from \textsc{Scheduling}.  The
optimal unaggregated routing tree problem for equal energy nodes is
similar to an NP-complete problem that has been studied in the network
design literature by the name of \textsc{Capacitated Spanning Tree}
\cite{garey79computers}.  In \textsc{Capacitated Spanning Tree}
problem we are given a graph $G$ with edge weights, a bound $K$ and a
root node.  The problem is to find a minimum weight spanning tree such
that every subtree that is hanging from the root node contains fewer
than $K$ vertices.

Now that we have seen that the optimal routing tree problem is 
NP-complete we would like to investigate the possibility of approximation
algorithms for the problem.  To evaluate the effectiveness of these
approximation algorithms, it is useful to have an upper bound on the
optimal solution.  In the following section, we formulate the routing
problem as a linear program which leads us to an upper bound.

\subsection{An Integer Programming Formulation}
\label{sec-integer-programming}
In this section, we formulate the decision problem as an integer
linear program.  To do this, we define two sets of \emph{integer}
variables $x_{ij}$ and $f_{ij}$ for every edge $ij$ on the graph.  If
$j$ is the parent of $i$ in the routing tree, then $x_{ij} = 1$, else
$x_{ij} = 0$.  $f_{ij}$ is the data flow that $i$ sends to $j$ along
the edge $ij$ per unit time.  We are interested in knowing whether
there exists a routing tree with lifetime $T$.  In that case,
feasibility of the following integer linear program is equivalent to
the decision problem.
\begin{eqnarray}
  \sum_{j = 1}^N x_{ij} & = & 1, \;\;\; i = 2, \ldots N 
  \label{cond1}\\
  \sum_{j = 1}^N f_{ij}-\sum_{j = 2}^N f_{ji} & =  & 1, \;\;\; i = 2,
  \ldots N 
  \label{cond2}\\
  x_{ij} \leq f_{ij} & \leq & \frac{e_i}{T}x_{ij}
  \label{cond3}\\
  x_{ij} & \geq & 0 
  \label{cond4}\\
  f_{ij} & \geq & 0 
  \label{cond5}
\end{eqnarray}
The condition (\ref{cond1}) ensures that every node has exactly one
parent, i.e. the set of edges for which $x_{ij} = 1$ define a tree.
Condition (\ref{cond2}) is a flow conservation condition.  The outflow
from a node is exactly one unit larger than inflow.  Condition
(\ref{cond3}) enforces two things.  First, if $j$ is $i$'s parent,
then $j$ must send at least one unit of flow to its parent.  Also if
$j$ is not $i$'s parent, then flow along the edge $ij$ is zero.
Second, it makes sure that the outflow from node $i$ respects the
energy constraint of node $i$.  Conditions (\ref{cond4}) and
(\ref{cond5}) are usual positivity conditions.

Of course, finding a feasible solution to the integer program is still
NP-complete, but it gives us a way to find an upper bound for the
problem by relaxing the integrality conditions.  Relaxing the variable
$x_{ij}$ to be any real number between 0 and 1 means that we allow
data to be forwarded to more than one node.  Thus the relaxed linear
program produces a solution which is no longer a tree.  In fact, we
shall show that the relaxed problem is equivalent to a maximum flow
problem.  With relaxation of integrality conditions, $f_{ij}$ can also
be arbitrary real numbers.  At this point the relaxed linear program
can be written down without the $x_{ij}$ variables at all.  The new
non-integer linear program is
\begin{eqnarray}
  \sum_{j = 1}^N f_{ij}-\sum_{j = 2}^N f_{ji} & =  & 1, \;\;\; i = 2,
  \ldots N 
  \label{cond6}\\
 \sum_{j=1}^N f_{ij} & \leq & \frac{e_i}{T}
  \label{cond7}\\
  f_{ij} & \geq & 0 
  \label{cond8}
\end{eqnarray}
If we define a new variable $y_{ij} \equiv T f_{ij}$, then the
decision problem can be written as a maximization problem:
\begin{eqnarray}
  \textrm{Maximize $T$, subject to} \nonumber \\
  \sum_{j = 1}^N y_{ij}-\sum_{j = 2}^N y_{ji} & =  & T, \;\;\; i = 2,
  \ldots N 
  \label{cond9}\\
 \sum_{j=1}^N y_{ij} & \leq & e_i
  \label{cond10}\\
  y_{ij} & \geq & 0 
  \label{cond11}
\end{eqnarray}
This linear program is equivalent to a maximum flow problem where each
node (except the root) has a node capacity $e_i$ and acts as a source
which sends a flow $T$ to the sink which is the root.  Using
conventional maximum flow algorithms\cite{ahuja93network}, this
problem can be solved in polynomial time.  Let us call the maximum
lifetime solution to this flow problem $T_\textrm{LP}$.  If the
optimum lifetime achievable for the original problem with tree routing
is $T_\textrm{OPT}$, then $T_\textrm{LP} \geq T_\textrm{OPT}$.  From
now on, we shall use $T_\textrm{LP}$ as an upper bound on
$T_\textrm{OPT}$.
\begin{figure}
  \includegraphics[width=0.45\textwidth]{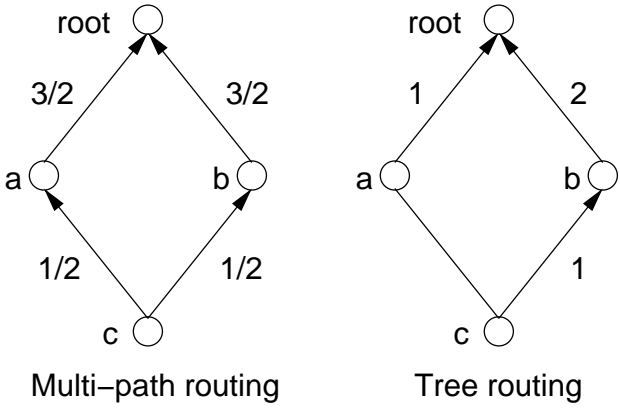}
\caption{Multipath routing and tree routing.  The numbers next to the
edges show the flow through that edge.}
\label{fig-flow}
\end{figure}

For an example graph where $T_\textrm{LP} > T_\textrm{OPT}$, consider
the topology shown in Fig. \ref{fig-flow}.  We assume that all nodes
have energy $1$.  In that case, the figure at right shows an optimal
routing tree with lifetime $1/2$.  The numbers next to the edges
denote the amount of data flow along that edge.  On the other hand in
a maximum flow solution, we can make the bottom node route its data
with probability $1/2$ along one link and with probability $1/2$ along
the other link.  In this case the data flow is more equally
distributed and the system lifetime is $2/3$.  We call this type of
routing where outflow from a node can reach the root via multiple
paths as \emph{multipath} routing.

Although sensor databases use tree based routing, it is not essential
to their functioning; the only requirement for correct functioning is
that there should be no duplication or loss of data packets.  Although
multipath routing seems to yield significant energy savings, there are
some hidden costs to it.  For example if data is transmitted
probabilistically to multiple nodes, all the receiving nodes need to
be listening in even if the packet is not destined for them.  As we
have mentioned, even idle listening consumes significant amount of
energy.  Thus the benefits of randomized multipath routing depends on
the complex interaction between MAC and network protocols.  In this
work we restrict ourselves to tree routing only.


\subsection{Energy Conserving Routing Tree (ECRT)}
Now we give a heuristic algorithm for computing an approximate routing
tree whose basic idea is very similar to Prim's algorithm for the
minimum spanning tree.  We initialize the routing tree to be the root
at first and grow the tree by adding edges one by one until the tree
spans all the vertices.  At every step we can determine the lifetime
of the tree constructed so far by calculating the amount of data each
node will be required to forward.  To decide which edge to add, we add
the edge that will maximize the lifetime of the resulting new tree.  If
there are multiple edges which result in the same lifetime, we choose
the edge which will add the node with maximum energy.  We call the
lifetime of the tree achieved by this algorithm $T_\textrm{ECRT}$.
The algorithm is formally described as \textsc{ECRT} in algorithm
\ref{algo-greedy-tree}.

\begin{algorithm}
\caption{ECRT($G(V,E), e$)}
\label{algo-greedy-tree}
\begin{algorithmic}[1]
\STATE Initialize tree $\mathcal{T}$ to contain the single node root.
\WHILE {Not all nodes are in $\mathcal{T}$}

\STATE Find lifetime $T$ of the tree $\mathcal{T}$.

\STATE Find the set of nodes $\mathcal{N}$ adjacent to $\mathcal{T}$.

\STATE Add node $v \in \mathcal{N}$ which reduces $T$ by the least amount
\ENDWHILE
\STATE Output $\mathcal{T}$.
\end{algorithmic}
\end{algorithm}

The algorithm \textsc{ECRT} has $N$ stages and in each stage a new
edge is added.  To find the optimal edge to add, we need to check
$\mathcal{O}(E)$ edges and for each edge, we need to check the
lifetime of the resulting tree.  If the diameter of the graph is $d$,
then to determine the life of the new tree takes no more than
$\mathcal{O}(d)$ time.  Thus \textsc{ECRT} runs in $\mathcal{O}(NEd)$
time.  This algorithm is not a constant factor approximation algorithm
and we state it formally as the following lemma.  In the interest of
brevity, the proof \cite{buragohain04power} is omitted.

\begin{lemma}
In the worst case the following lower bound exists for the performance
ratio of the ECRT algorithm
\begin{displaymath}
  \frac{T_\textrm{OPT}}{T_\textrm{ECRT}} = 
  \Omega(\log  N)
\end{displaymath}
\end{lemma}
In other words, there exists instances of connectivity graphs where
the optimal lifetime is better than the lifetime achieved by
ECRT by a factor of $\log N$ or larger.

Now that we have seen that ECRT is not an optimal
algorithm for the routing problem, we ask the question if there is
some optimization that can improve is performance.  With this aim, we
present an algorithm which optimizes a pre-existing routing tree.

\subsection{Local Optimization}
Consider an optimal routing tree.  If any single node in the tree
switches its parent to another node, the resulting tree will have a
lifetime less than or equal to the optimal tree.  Thus for the optimal
solution, the choice of parent for every node is optimal.  We now
define a routing tree to be \emph{locally optimal}, if the lifetime of
the tree can not be improved by switching the parent of \emph{any
single} node.  Note that the optimal tree is always locally optimal,
but the converse need not be true.

This criterion immediately suggests an approximation algorithm for
finding a locally optimal tree which we call LOCAL-OPT.  The algorithm
accepts as input an arbitrary routing tree and considers the nodes
sequentially in some particular order.  We try to see if the lifetime
of the tree will be improved by switching the parent of the current
node in consideration.  If such a switch is favorable, it is made;
otherwise we proceed to the next node.  We call this switch an
\emph{improvement step}.  We denote the lifetime achieved by this
algorithms as $T_\textrm{LO}$ and describe it formally as algorithm
\ref{algo-local-opt}.

\begin{algorithm}
\caption{LOCAL-OPT($G(V,E), \mathcal{T}, e$)}
\label{algo-local-opt}
\begin{algorithmic}[1]
\STATE done $\leftarrow$ FALSE
\WHILE {done = FALSE }
\STATE done $\leftarrow$ TRUE
\FORALL{$v \in V$}
\IF{switching the parent of $v$ improves $\mathcal{T}$}
\STATE switch parent of $v$ to improve $\mathcal{T}$
\STATE done $\leftarrow$ FALSE
\ENDIF	
\ENDFOR
\ENDWHILE
\STATE Output $\mathcal{T}$.
\end{algorithmic}
\end{algorithm}

This algorithm is not a constant factor approximation algorithm for
the unaggregated routing tree problem as shown by the following lemma.
In the interest of brevity, the proof \cite{buragohain04power} is
omitted.
\begin{lemma}
In the worst case the following lower bound exists for the performance
ratio of the LOCAL-OPT algorithm
\begin{displaymath}
  \frac{T_\textrm{OPT}}{T_\textrm{LO}} = 
  \Omega\left(\frac{\log  N}{\log\log N}\right)
\end{displaymath}
\end{lemma}
In simpler words, there exists an instance of a graph where the
optimal lifetime is better than the lifetime produced by LOCAL-OPT by
a factor of $\log N/\log\log N$ or more.  The LOCAL-OPT algorithm can
be used as a stand alone algorithm to find an approximate routing
tree, or it can be used as an additional optimization on the tree
produced by ECRT algorithm.


For performance analysis of LOCAL-OPT, consider a network with maximum
and minimum energies $e_\textrm{max}$ and $e_\textrm{min}$
respectively.  Then the maximum lifetime for \emph{any} routing tree
is $\frac{e_\textrm{max}}{\textrm{deg(root)}N}$ where deg(root) is the
degree of the root node.  Let us now look at the size of a local
improvement step: the smallest size of the improvement step is
\begin{displaymath}
\frac{e_\textrm{min}}{N-1}-\frac{e_\textrm{min}}{N} \approx \frac{e_\textrm{min}}{N^2}.
\end{displaymath}
Then the total number of improvement steps are bounded by
\begin{displaymath}
\frac{e_\textrm{max}}{e_\textrm{min}}\frac{N}{\textrm{deg(root)}}
\end{displaymath}
Thus LOCAL-OPT runs in time polynomial in
$e_\textrm{max}/e_\textrm{min}$ and $N$.

\section{Routing Tree for Partially Aggregated Queries}
\label{sec-partly-aggregated}
A partially aggregated query has a PSR size which lies between fully
aggregated queries and unaggregated queries.  Consider as an example a
histogram query with $\ell$ bins.  The PSR can be just a listing of
bins with their associated frequencies.  When the query reply starts
out at a leaf node the PSR consists of 1 bin which holds the value and
the corresponding frequency which is 1.  As more and more data points
are added, they begin to fall into distinct bins and the size of the
PSR increases.  But once the PSR size reaches $\ell$ bins, any new
data added falls into one of the older bins and hence the PSR size
reaches a constant.  So we model a partially aggregated query as
follows.  Every node produces one single unit of data every time
period.  As long as the total data inflow into a node is less than
$\ell$, the data remains conserved, i.e. if two PSRs reach a node,
then the new PSR size is just the sum of the two old PSR sizes and 1
(size of its own data).  If at any node the sum of the PSR sizes
exceed $\ell$, the PSR forwarded to the parent has size $\ell$ only.
For $\ell=N$, this problem reduces to the problem of unaggregated
queries.
	
The approximation algorithms used to solve the unaggregated routing
tree problem can be immediately adapted to solve this problem.  These
algorithm depend on being able to find the lifetime $T$ of a given
tree $\mathcal{T}$.  For the unaggregated case, we defined the
lifetime $T$ as the minimum of the ratio
$e_i/f_\textrm{out}(i)$ where $f_\textrm{out}(i)$ is the outflow from
node $i$.  For partially aggregated queries $f_\textrm{out}$ is bound
by $\ell$.  With this modification, both the ECRT and LOCAL-OPT
algorithms can be easily applied to this problem.

\section{Experimental Results} 
\label{sec-experimental}
\begin{figure}
  \includegraphics[width=0.6\textwidth]{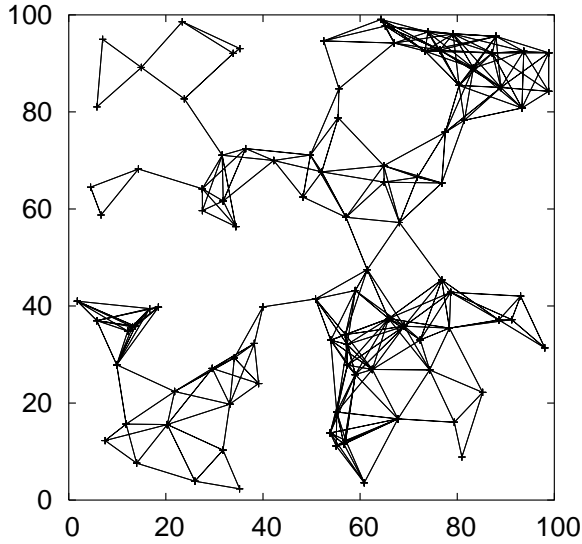}
\caption{An example connectivity graph for 100 nodes randomly placed
  in a $100\times 100$ area with radio range set to 15 ($r = 1.5$).}
\label{fig-graph}
\end{figure}

We evaluated our algorithms on simulated communication graph
topologies.   The simulation parameters are described below.
\begin{itemize}
  \item \emph{Node Distribution}: To generate the graph we randomly
place $N$ nodes in an area of size $d\times d$.  The node numbered 1
is arbitrarily chosen as the base station.
\item \emph{Connectivity}: We set the radio range to be $l_r$.  Since
our length units are arbitrary, we define a scaled radio range $r$ as
follows:
\begin{displaymath}
  r \equiv \frac{l_r}{l_0},\;\; l_0 \equiv
  \left(\frac{d^2}{N}\right)^{1/2}.
\end{displaymath}
One can think of $l_0$ is the average separation between nodes.  Thus
for a given placement of nodes, the communication graph consists of
nodes as vertices and edges between nodes which are within range of
each other.  We expect that for radio range $l_r \lessapprox l_0$, the
graph will be disconnected.  In practice the approximate connectivity
threshold turns out to be $r = l_r/l_0 \lessapprox 1.5$.  A
representative connectivity graph with $r = 1.5$ is shown in 
Fig. \ref{fig-graph}.
\item \emph{Energy Distribution}: The energy levels for nodes are
chosen randomly from a uniform distribution between $e_\textrm{max} $,
the maximum and  $e_\textrm{min}$, the minimum.  The range of
possible values of energy are controlled by tuning the \emph{energy
ratio} $\alpha$ defined by
\begin{displaymath}
  \alpha = \frac{e_{\textrm{max}}}{e_{\textrm{min}}}.
\end{displaymath}
We chose $e_\textrm{max}$ and $e_\textrm{min}$ such that
the average is 1000.  For example when $\alpha = 1$, all nodes are
assigned energy 1000, while for $\alpha = 2$, nodes are assigned
random energy values between 667 and 1333.

\item \emph{Energy Cost}: we assume that transmitting one unit of
data costs 1 unit of energy, while receiving the same amount of data
costs 0.5 units, i.e. $c_r = 0.5$.
\end{itemize}

Currently the sensor databases use shortest path routing with path
length equal to number of hops.  We call this strategy
\textsc{MIN-HOP} and compare our algorithms \textsc{ECRT}, and
\textsc{LOCAL-OPT} to it.  Note that the routing tree produced by the
MIN-HOP algorithm is same as the breadth-first-search (BFS) tree for
the graph.

\subsection{System Lifetime for Fully Aggregated Queries}

\begin{figure}[ht]
  \includegraphics[width=0.45\textwidth]{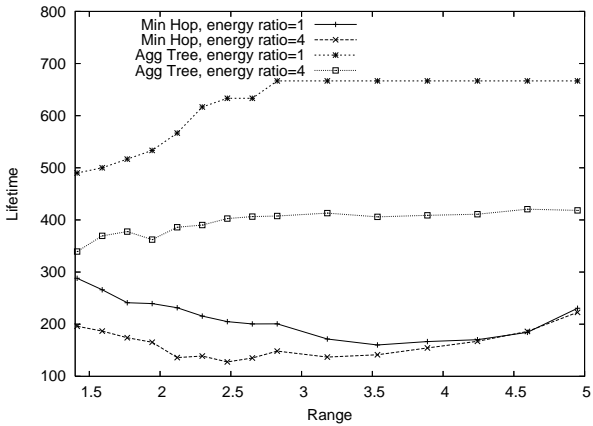}
\caption{System lifetime plotted against scaled range ($r$).  There
  are $N = 50$ nodes.}
\label{fig-mdst-life-vs-range}
\end{figure}
In Fig. \ref{fig-mdst-life-vs-range} we plot the performance of the
AGGREGATED-TREE algorithm vs the MIN-HOP algorithm.  We used two
different energy distributions : $\alpha = 1$, i.e. $e_\textrm{max} =
e_\textrm{min} = 1000$ and $\alpha =4$ with $e_\textrm{max} = 1600,
e_\textrm{min} = 400$.

We note that AGGREGATED-TREE significantly outperforms the MIN-HOP
for all values of $r$.  But apart from that there are several
interesting features to be seen in Fig. \ref{fig-mdst-life-vs-range}.
As the range $r$ is raised from 1.5 to 3, the performance of MIN-HOP
declines while the performance of AGGREGATED-TREE improves.  With
increasing range, the graph connectivity increases and MIN-HOP tree
becomes bushier.  Thus every node acquires more neighbors and MIN-HOP
tree lifetime decreases.  On the other hand, with more connectivity,
AGGREGATED-TREE is able to reduce the maximum degree and thus perform
better and better.  After $r > 3.5$, the MIN-HOP lifetime begins to
rise again because now most of the nodes are being able to connect to
the root directly.  In contrast, the AGGREGATED-TREE lifetime
saturates.  The optimal lifetime tree for energy ratio $\alpha = 1$ is
limited by the maximum degree which can not be smaller than 2.  For
degree 2, the lifetime is $1000/(1+0.5) = 667$ and the graph shows
that for $r\geq 3.0$, this lifetime is achieved.  For energy ratio
$\alpha = 4$, the lifetime is limited by the minimum energy node which
has energy $e_\textrm{min} = 400$.  A lifetime of 400 can be achieved
if the minimum energy node is a leaf and for $r \geq 3.0$, this is
achieved (eqn. \ref{eqn-upper-bound}).

\subsection{System Lifetime for Unaggregated Queries}
\begin{figure}[ht]
  \includegraphics[width=0.45\textwidth]{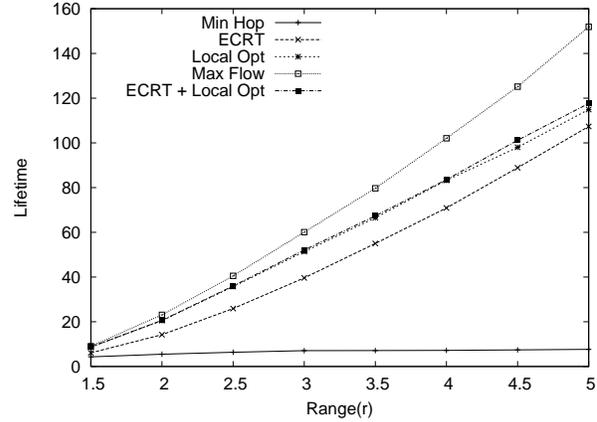}
\caption{System lifetime plotted against scaled range ($r$).  There
  are $N = 400$ nodes with all equal energy ($\alpha = 1$) of 1000 units.}
\label{fig-life-vs-range}
\end{figure}
In Fig. \ref{fig-life-vs-range} we plot the performance of ECRT,
LOCAL-OPT, and MIN-HOP algorithms as a function of radio range.  As an
upper limit, we also plot $T_\textrm{LP}$ which is equivalent to a
maximum flow on the graph.  Naturally the nodes closest to the root
node are the most heavily loaded.  As we increase the range $r$, the
graph begins to become more and more connected and the number of nodes
next to the root begins to rise; hence the lifetime increases.  As
expected, MIN-HOP performs the worst because it is not sensitive to
energy.  The LOCAL-OPT algorithm was initialized with a MIN-HOP tree
and it performs better than ECRT.  We also plotted the effect of
optimizing the solution of the ECRT algorithm by using LOCAL-OPT on
it.  The optimized solution is only marginally better than LOCAL-OPT
by itself.

\begin{figure}[ht]
  \includegraphics[width=0.45\textwidth]{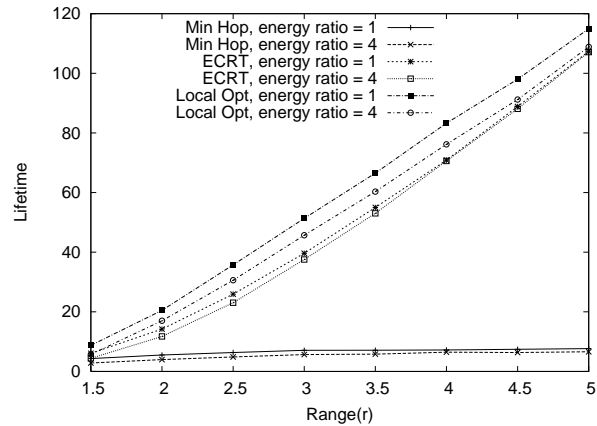}
\caption{Effect of changing the distribution of energy values on
  system lifetime ($N=400$).}
\label{fig-life-random}
\end{figure}

The effect of changing the distribution of energy values is shown in
Fig. \ref{fig-life-random}.  We varied the energy ratio $\alpha =
e_\textrm{max}/e_\textrm{min}$ from 1 to 4.  The effect on the average
system lifetime was minimal.  


\subsection{Effect of Network Size on Lifetime for Unaggregated Queries}
\begin{figure}[htb]
  \includegraphics[width=0.45\textwidth]{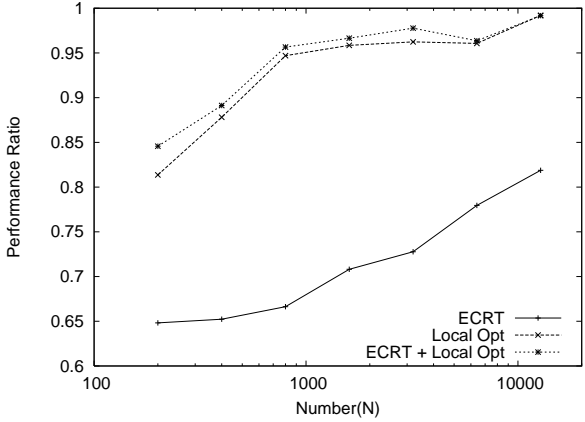}
\caption{Performance ratio achieved by ECRT and LOCAL-OPT
routing algorithms ($r = 3.0$).}
\label{fig-flow-greedy-ratio}
\end{figure}
Consider a network with all nodes having equal energy.  For any tree
algorithm, the best performance is achieved when the sizes of the
subtrees that are rooted at the immediate neighbors of the root node
are equal.  The chance of doing this increases as the number of nodes
and hence number of paths in the network increases.  So we expect that
with increasing $N$, our approximation algorithms will perform close
to optimal.  Quantitatively, we define a quantity called
\emph{performance ratio} as follows:
\begin{displaymath}
  \textrm{Performance ratio} =
  \frac{T_\textrm{APPROX}}{T_\textrm{LP}}.
\end{displaymath}
Closer this ratio is to 1, better the algorithm.  In
Fig. \ref{fig-flow-greedy-ratio}, we plot the performance ratio of the
ECRT and LOCAL-OPT algorithms as a function of $N$.  The graph clearly
demonstrates that both algorithms approach the optimal solution as $N$
increases, but LOCAL-OPT is much quicker in convergence.  Optimization
of the ECRT solution by the LOCAL-OPT algorithm yields marginally
better results.

\subsection{System Lifetime for Partially Aggregated Queries}
\begin{figure}[ht]
  \includegraphics[width=0.45\textwidth]{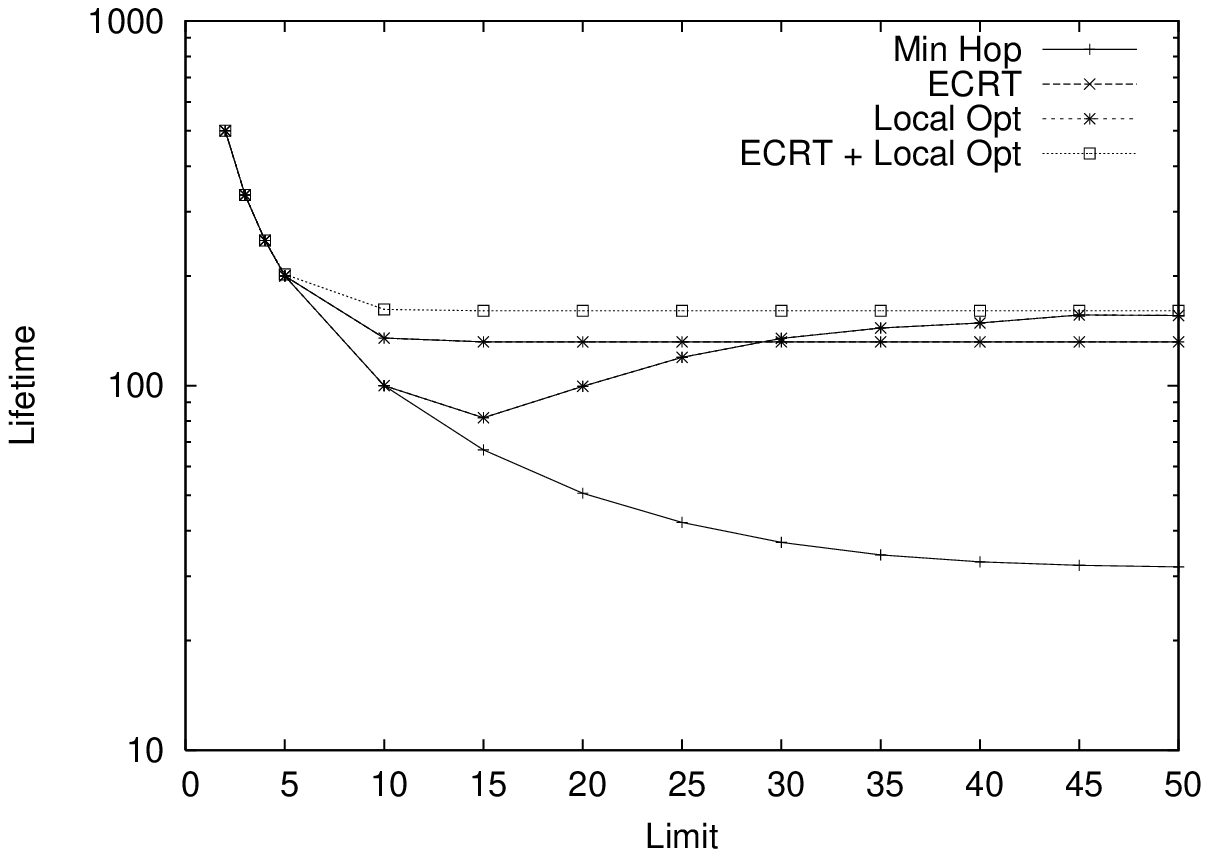}
  \includegraphics[width=0.45\textwidth]{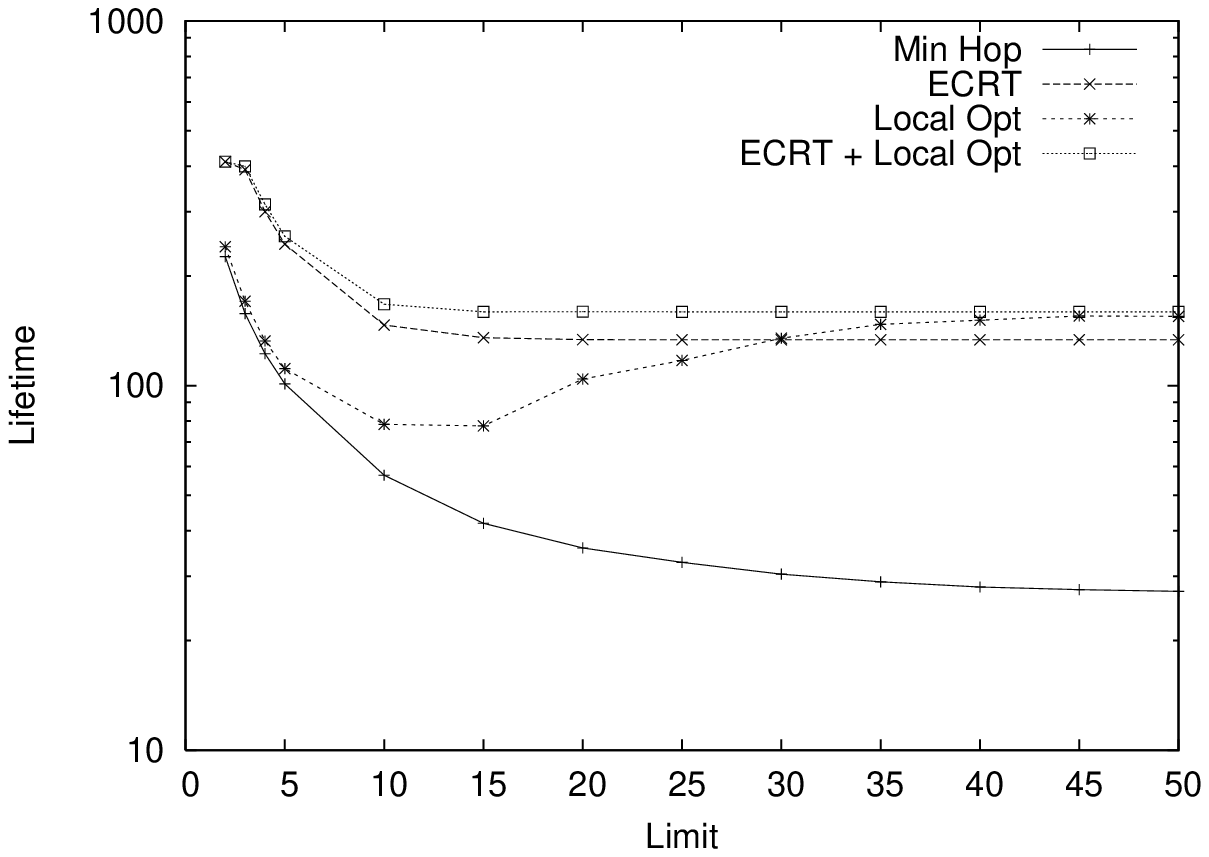}
  \caption{System lifetime plotted as function of data flow limit.
  The top graph shows lifetime for equal energy levels ($\alpha=1$),
  while the bottom graph shows lifetime for unequal energy levels
  ($\alpha=4$).  The total number of nodes $N$ is 100 and range is $r=3.0$.}
\label{fig-life-vs-limit}
\end{figure}
Recall that for partially aggregated queries, the message size (or the
PSR size) increases from leaves towards the root, but it can not
exceed a limit which we call $\ell$.  We plot the system lifetime as a
function of $\ell$ in Fig. \ref{fig-life-vs-limit}.  For large $\ell
(\ell \geq N)$, this problem becomes identical to the unaggregated
query problem.  The upper graph shows the results for equal energy
case ($\alpha = 1$), while the lower graph shows the results for
unequal energies ($\alpha=4$).  

As expected, we see that LOCAL-OPT and ECRT algorithms perform
significantly better than the \textsc{MIN-HOP} algorithm for most
values of $\ell$.  For small values of $\ell$, being able to
distribute load uniformly across sensors is no longer crucial and thus
all the algorithms perform similarly.  Although for large $\ell$,
LOCAL-OPT outshines ECRT, for intermediate values of $\ell$, ECRT does
better.  We conjecture that for small values of $\ell$, small local
changes in the routing tree do not lead to improvements in lifetime
and thus LOCAL-OPT stops improvement steps too early.  The best
results are achieved when we optimize the output of the ECRT algorithm
by using it as the input to LOCAL-OPT.


\section{Conclusion}
\label{sec-conclusion}
We have argued that the problem of energy efficient routing in sensor
databases is intimately connected with the type of query that the
database is expected to answer.  For many types of queries, we have
shown that the problem of efficient tree routing is NP-complete.  For
fully aggregated queries, we have given a constant factor
approximation algorithm; for other problems we have given heuristic
algorithms which exhibit excellent performance in practice and for
large networks approach the optimal solution.


\bibliographystyle{IEEE}

\end{document}